# Leveraging Multi-Source Textural UGC for Neighbourhood Housing Quality Assessment: A GPT-Enhanced Framework

Qiyuan Hong[*1], Huimin Zhao[†1] and Ying Long[‡12]

[1]School of Architecture, Tsinghua University
[2]School of Architecture and Hang Lung Center for Real Estate, Key Laboratory of Eco Planning & Green Building, Ministry of Education, Tsinghua University, Beijing, China

**Summary**

This study leverages GPT-4o to assess neighbourhood housing quality using multi-source textural user-generated content (UGC) from Dianping, Weibo, and the Government Message Board. The analysis involves filtering relevant texts, extracting structured evaluation units, and conducting sentiment scoring. A refined housing quality assessment system with 46 indicators across 11 categories was developed, highlighting an objective-subjective method gap and platform-specific differences in focus. GPT-4o outperformed rule-based and BERT models, achieving 92.5% accuracy in fine-tuned settings. The findings underscore the value of integrating UGC and GPT-driven analysis for scalable, resident-centric urban assessments, offering practical insights for policymakers and urban planners.

## 1. Introduction

Neighbourhood housing quality (HQ) directly affects residents' physical and mental health (Hadley-Ives et al., 2000), making its assessment crucial for urban planning and renewal. Traditional methods, such as interviews and field surveys, are often labor-intensive and spatially and temporally limited (Zhao and Long, 2022). The advent of emerging technologies, including remote sensing and unmanned aerial systems, has facilitated more efficient and quantitative evaluations of environmental quality (Grubesic et al., 2018). However, these approaches emphasize objective metrics while overlooking residents' subjective perceptions, which are crucial to neighbourhood health and well-being (Kovacs-Györi et al., 2020).

User-generated content (UGC) on social media provides a novel avenue to capture residents' needs and preferences. UGC provides both subjective satisfaction insights and objective feedback on neighbourhood conditions. Recent studies have leveraged UGC from housing platforms like Airbnb and Zigbang and social media like Twitter, employing NLP techniques like BERT to extract semantic information and sentiments to assess neighbourhood perceptions (Kweon and Lee, 2023; Wang et al., 2023; Hollander et al., 2016).

However, existing methods struggle to fully extract insights from unstructured text, as they mostly rely on quantitative statistics, sentiment classification, or predefined labels, limiting analytical depth (Lore et al., 2024). Additionally, the lack of multi-platform integration ignores topic variations across sources, potentially biasing assessments (Alipour et al., 2023).

---

[*] hongqy24@mails.tsinghua.edu.cn
[†] zhaohm21@mails.tsinghua.edu.cn
[‡] ylong@tsinghua.edu.cn

To address these challenges, we introduces GPT models, which offer advanced contextual understanding and noise handling without extensive labeled datasets. This study examines GPT models' feasibility in assessing neighbourhood HQ from multi-platform UGC. By extracting content and sentiment, it develops an HQ evaluation framework that prioritizes residents' subjective perceptions. A case study in Beijing demonstrates the model's applicability.

## 2. Methodology

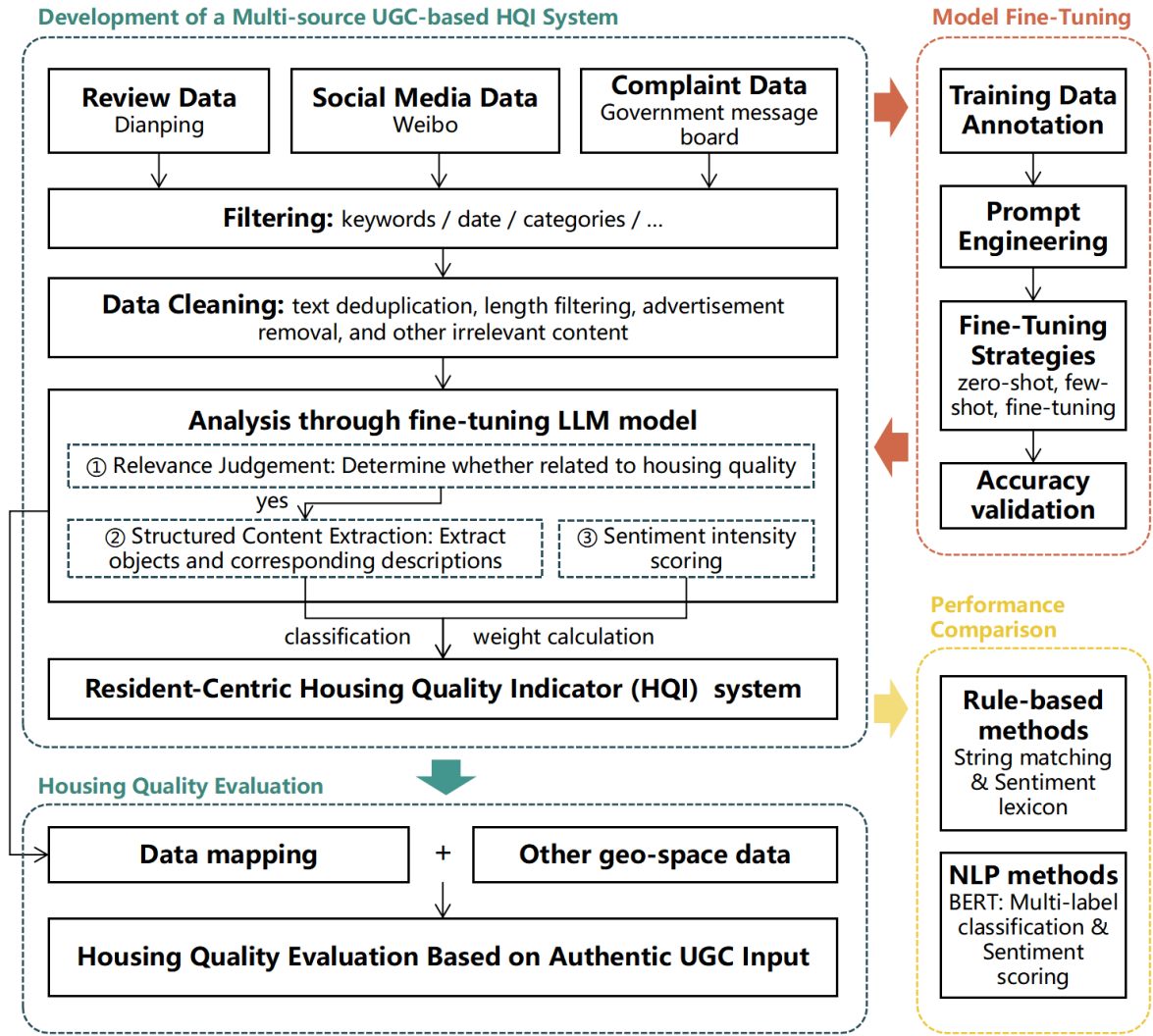

**Figure 1** Research Framework.

### 2.1. Data Processing and Analysis

Shown as **Figure 1**, textual data were sourced from three platforms: Dianping (reviews), Weibo (social media), and the Government Message Board (complaints). After filtering and cleaning, GPT-4o was used for three tasks: Relevance Judgement; Structured Content Extraction - forming structured units; and Sentiment Scoring - assigning sentiment scores ($S_i$) on a five-point scale ranging from strongly negative (-2) to strongly positive (2).

A randomly selected sample of 1,000 entries was manually annotated for model fine-tuning. Referencing established HQ indicators (MOHURD, 2024; UK Homes and Communities Agency, 2008), we summarized the extracted objects into categories and the content descriptors into indicators.

## 2.2. Accuracy Validation of GPT models and Baseline Comparisons

The feasibility of GPT models for assessment were validated on 200 samples. First, we compared zero-shot, few-shot (trained on 10 samples), and fine-tuned models to determine the effect of different training strategies on accuracy.

Second, we compared GPT's generative approach with two classification-based methods: a rule-based String Search method and a widely used NLP model BERT. For the rule-based method, we applied predefined categories and the HOWNET lexicon for sentiment analysis. BERT was trained for multi-label classification and sentiment prediction.

## 2.3. Construction of the Housing Quality Evaluation System

A housing quality evaluation system was developed, incorporating indicator weights based on sentiment intensity and frequency, calculated as:

$$F'_i = \log(F_i + 1) \tag{1}$$

$$W_i = \frac{I_i \cdot F_i}{\sum_{i=1}^{n}(I_i \cdot F_i)} \tag{2}$$

Where:
$W_i$: Weight of indicator $i$
$I_i$: Importance of $i$, $I_i=|S_i|$
$F_i$: Frequency of $i$
$F'_i$: Logarithmic frequency
$n$: Total number of indicators

## 2.4. Evaluation of Residential Communities in Beijing

We used multi-source data to evaluate communities in Beijing. Baidu AOI and POI data helped determine spatial boundaries, and housing quality assessments were mapped to specific communities. The scoring range was from 1 to 5, calculated as:

$$Total\ Score = \sum W_i\ |S_i + 3| \tag{3}$$

With unmentioned $i$, $S_i$ is scored as neutral.

## 3. Results

### 3.1. Comments Analysis and HQI System Construction

A total of 48,000, 409,000, and 210,000 entries from the platforms, posted during 2023–2024, were collected. These data were processed into 275,000 structured units, each consisting of an evaluation object and a corresponding content description.

As shown in **Figure 2**, Quantity and coverage of parking spaces (Indicator 4.1) had the highest frequency (22,943 units), while Accessible Parking Spaces (Indicator 10.4) had the lowest (38 units). Indicator 4.4 exhibited the strongest sentiment intensity, whereas Indicator 1.3 had the weakest. Based on these findings, we identified 46 indicators across 11 categories and determined their respective weights.

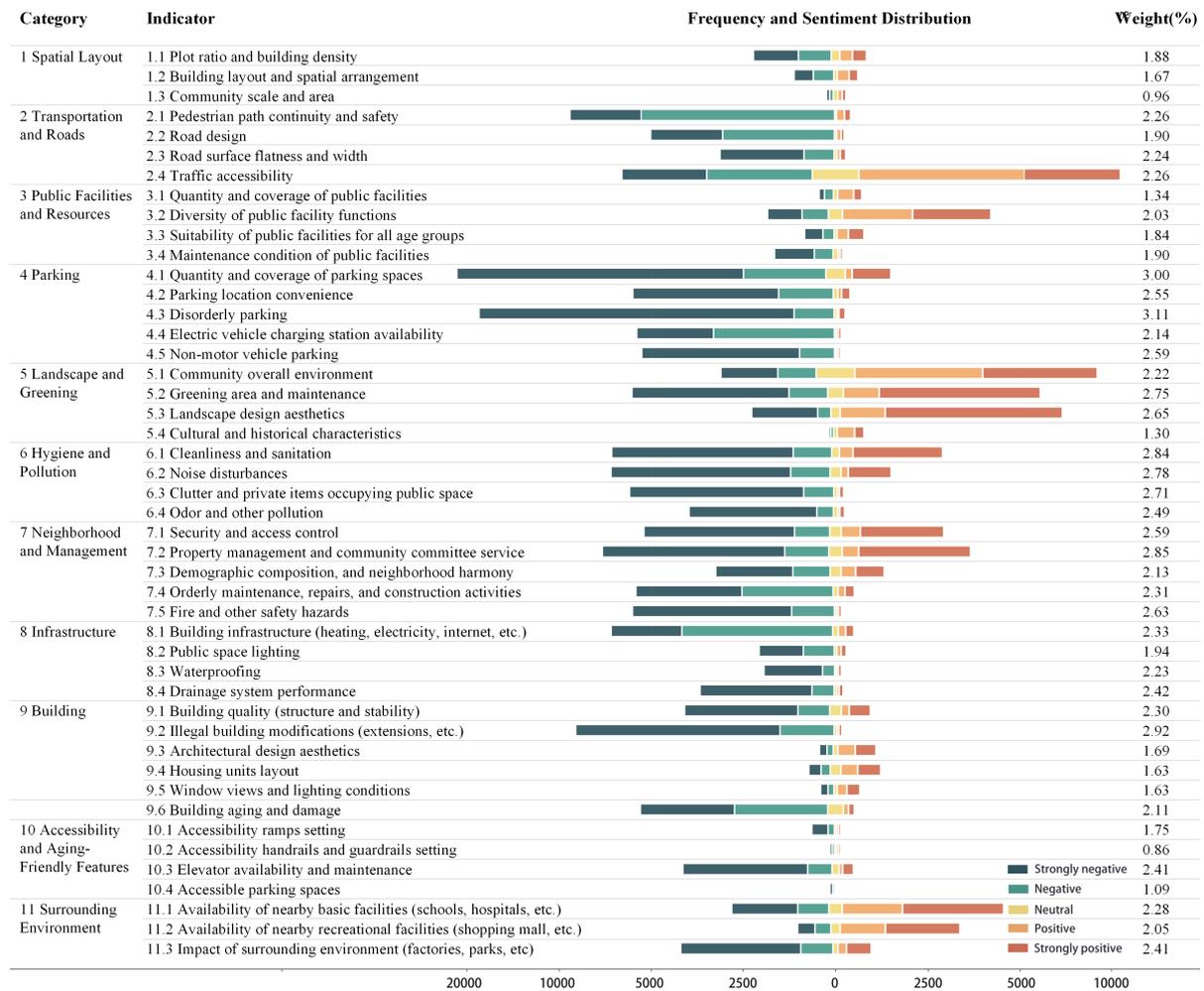

**Figure 2** Construction of the HQI system.

Additionally, the study identified platform-specific differences in HQ discussions. Dianping users emphasized Transportation and Roads (19%) and Community Overall Environment (15%); Weibo users focused on Neighbourhood and Management (21%); while Parking issues (22%) dominated Government Message Board complaints.

## 3.2. Accuracy Assessment of GPT and Baseline Methods

The GPT-4o model demonstrated strong performance, achieving 52.1% accuracy in zero-shot, 87% in few-shot, and 92.5% in fine-tuned modes. Comparatively, rule-based and NLP methods underperformed: String Search achieved(15.9%) suffered high false positives; sentiment dictionary-based methods reached 25%. BERT achieved 23.4% for object classification but only 5.2% for content classification. BERT achieved 88% accuracy in sentiment intensity classification.

## 3.3. Evaluation Results for Beijing

Within Beijing's Fifth Ring Road, data covered 2,038 out of 6,198 residential communities (**Figure 3**). The average HQ score was 3.40 out of 5. Indicators in Public Facilities and Resources performed the best, while the Parking category showed the weakest performance.

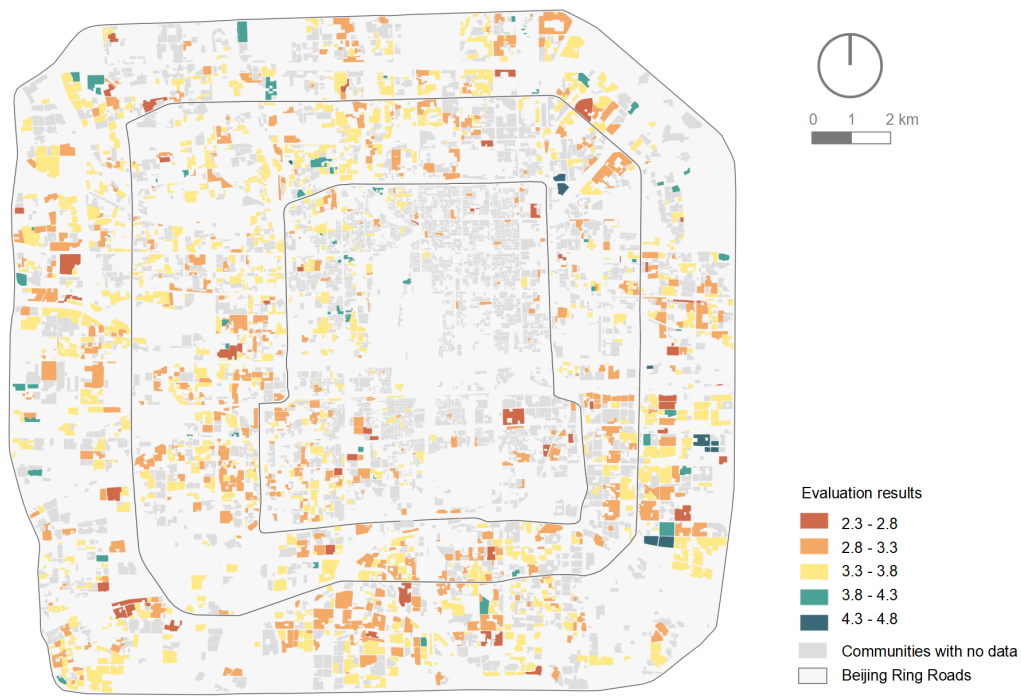

**Figure 3** Beijing Evaluation results.

## 4. Conclusions and Discussion

This study proposes a GPT-based approach to assess neighbourhood housing quality (HQ) from multi-source UGC. The primary contribution is a large-scale evaluation framework emphasizing residents' subjective perceptions. GPT-4o effectively judged topic relevance, extracted objects and contents, and assessed sentiment. Notably, it demonstrated strong generalization with minimal labeled data, outperforming rule-based and NLP models. Few-shot and fine-tuned models achieved 87% and 92.5% accuracy, significantly outperforming string search (15.9%) and BERT (23.4%&88%) with the same training set.

This study develops a refined HQ assessment framework with 46 indicators across 11 categories, incorporating often-overlooked aspects such as EV charging infrastructure. It also identifies platform-specific discussion differences, underscoring the need for multi-source data integration. An evaluation of 2,038 Beijing communities yielded an average HQ score of 3.40. Comparison with an objective HQ assessment (Zhao and Long, 2022) showed a weak correlation (0.15), highlighting the gap between objective and subjective measures. These findings reinforce the necessity of resident-driven HQ insights, offering valuable guidance for policymakers and urban planners.

Despite its strengths, this study has limitations. The indicator system was developed based on Chinese UGC data, restricting direct applicability to other regions. The analysis relies solely on text, excluding visual and spatial data that could enrich HQ assessments. Future research should integrate multimodal data, including images and geospatial information, to enhance framework robustness and validity.

**Biographies**

Qiyuan Hong is now a PhD student in the School of Architecture at Tsinghua University in China. Her research primarily revolves around urban science, with a particular focus on neighbourhood and communities, living environment quality, and planning support techniques.

Huimin Zhao is currently a PhD student at the School of Architecture at Tsinghua University in China. Her research interests are deeply rooted in the field of urban science, with a primary focus on housing vacancy and neighbourhood environmental quality.

Ying Long, PhD, is now a tenured professor in the School of Architecture, Tsinghua University. His research focuses on urban science, including applied urban modeling, urban big data analytics and visualization, data augmented design, and future cities. He has published almost two hundred papers and led over twenty research/planning projects.